\documentclass[prb,footinbib,amsmath,amssymb,twocolumn,floatfix,showpacs]{revtex4}

\usepackage{color}
\usepackage{hyperref}
\usepackage[pdftex]{graphicx}

\newcommand{\bs}{\;\;\;\;\;}
\newcommand{\sms}{\;\;}
\newcommand{\ve}{\mathbf}

\newcommand{\D}{{\rm d}}

\begin{document}

\title{Edge states and enhanced spin-orbit interaction at graphene/graphane interfaces}
\author{Manuel J. Schmidt}
\author{Daniel Loss}
\affiliation{Department of Physics, University of Basel, Klingelbergstrasse 82, 4056 Basel, Switzerland}
\date{\today}
\pacs{72.80.Vp, 75.70.Tj, 73.43.-f}

\begin{abstract}
We study interfaces between graphene and graphane. If the interface is oriented along a zigzag direction, edge states are found which exhibit a strong amplification of effects related to the spin-orbit interaction. The enhanced spin splitting of the edge states allows a conversion between valley polarization and spin polarization at temperatures near one Kelvin. We show that these edge states give rise to quantum spin and/or valley Hall effects.
\end{abstract}

\maketitle

Two-dimensional electronic systems in which the important physics is taking place solely at the edges have received considerable interest during the last decades, starting with the discovery of the quantum Hall effect (QHE) \cite{qhe}. In a strong magnetic field, gapless transport channels exist only at the edges of the system and these give rise to a remarkably well quantized conductance. For a long time, the QHE was the only known effect of this type. Recently, however, gapless modes have also been predicted \cite{kane_qshe,zhang_topological_insulator} and observed \cite{koenig_qshe,hasan} on the surface of topological insulators without the need of magnetic fields. The first system in which these states have been investigated is graphene \cite{kane_qshe,graphene1,graphene_rmp}. Here, spin-orbit interaction (SOI) plays the role of the magnetic field in that it opens a bulk energy gap. Spin-polarized transport is confined to the graphene edges which are oriented along the zigzag direction. It turned out, however, that the SOI in graphene is rather weak \cite{spin_orbit_macdonald}. Furthermore, it is difficult to actually construct clean structural edges of graphene, although some experimental progress has been reported \cite{edge_fabrication1}.

\begin{figure}[!h]
\centering
\includegraphics[width=200pt]{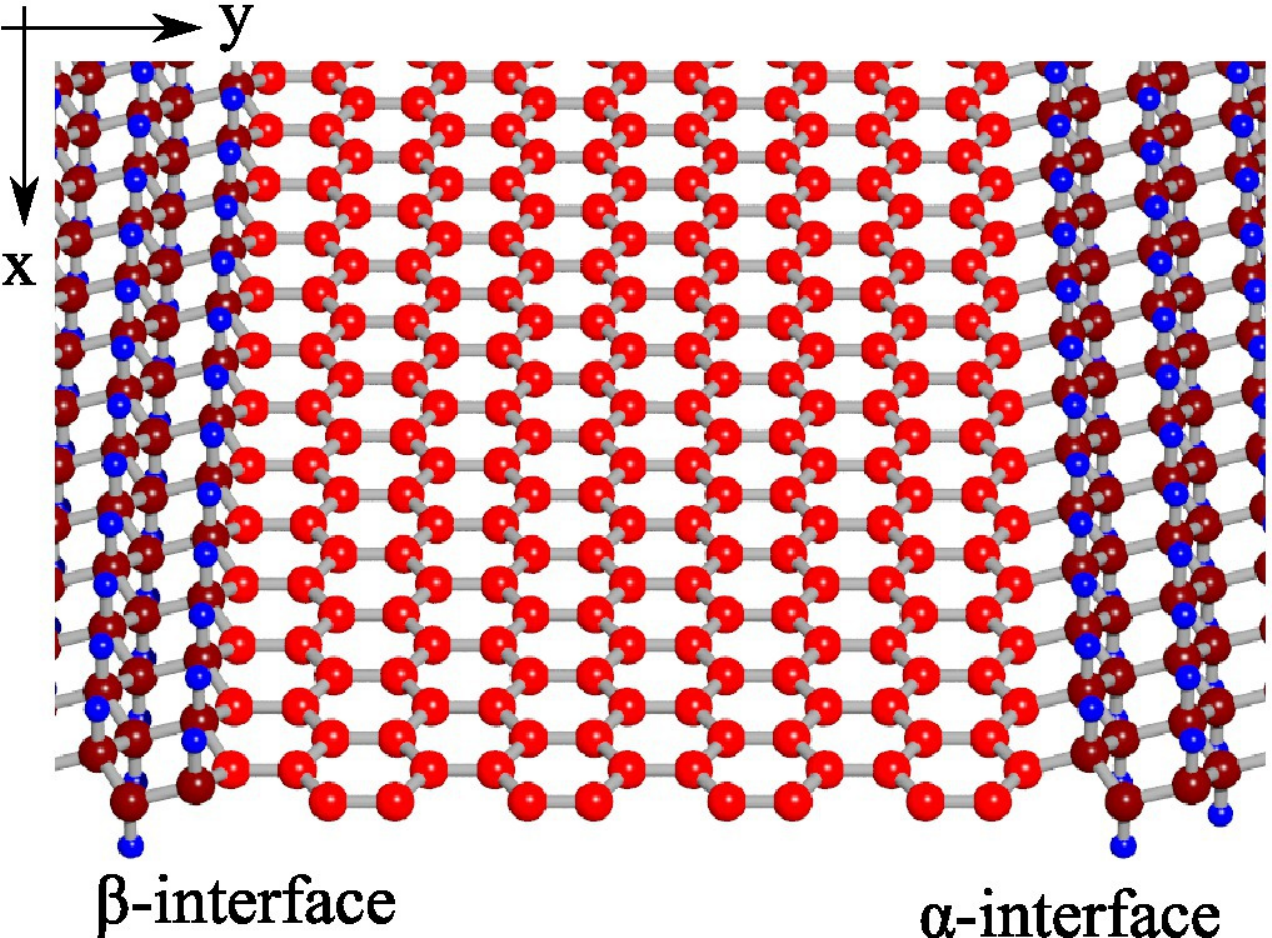}
\caption{(color-online) $\alpha$- and $\beta$-interfaces in a graphene/graphane heterostructure. The large (red) spheres represent the carbon atoms and the small (blue) spheres the hydrogen atoms. The brighter spheres correspond to the graphene region while the darker spheres correspond to graphane.}
\label{fig_alpha_beta_interface}
\end{figure}

Recently, a new possibility for the construction of very clean graphene terminations has been proposed \cite{graphane_terminated_graphene}: rather than structurally cutting graphene in order to create nanostructures, it has been proposed to hydrogenate it locally. Thereby, the $\pi$-band is removed wherever graphene is transformed into graphane \cite{graphane_theor_1,graphane_exp_1} and an {\it effective edge} is created for the $\pi$-electrons. By this technique of local hydrogenation, not only ribbons with high quality edges of the usual zigzag type (we call them $\alpha$-edges, henceforth) are possible; also bearded edges $(\beta$) are within reach (see Fig. \ref{fig_alpha_beta_interface}).

We study such graphene/graphane (GG) interfaces of $\alpha$- and $\beta$-type and find edge states which are exponentially localized at the interface. However, instead of being energetically nearly flat, like in the case of structural graphene edges, the edge states at GG interfaces have a considerable dispersion, the amplitude of which is largely determined by the energy of the hydrogen 1s-orbital. These edge states enable a quantum spin-valley Hall effect (QSVHE). Moreover, the SOI of these edge states is strongly enhanced due to the proximity of graphane. As an application we propose a device which is capable of converting valley polarizations into spin polarizations.

{\it Edge states.} Henceforth we assume that all edges and interfaces are oriented along a zigzag direction. Due to the two-atomic unit cell of a hexagonal lattice, there are in principle two different types of boundary conditions - $\alpha$-type and $\beta$-type (see Fig. \ref{fig_alpha_beta_interface}). The reason why a $\beta$-edge is seldom considered is that it is highly unstable against structural recombinations \cite{edge_stability}. Indeed, we are not aware of any experimental observation of a stable $\beta$-edge in graphene \cite{klein_footnote}. However, if we do not require the graphene flake to be cut in order to generate a $\beta$-edge, but only require the hydrogen deposition on graphene to start on a certain sublattice while the hexagonal structure remains intact, it makes not much of a difference whether $\alpha$- or $\beta$-interfaces are to be created.

Before we turn to a more elaborate modeling of GG interfaces and the related edge states which respects the $\pi$-band and the $\sigma$-band, we would like to start with a qualitative discussion on the basis of a very simplified model which, as it turns out, captures most of the important concepts. We begin with a nearest neighbor tight-binding model for the $\pi$-electrons in graphene
$H_G = t\sum_{\left<\ve r,\ve r'\right>} c_{\ve r}^\dagger c_{\ve r'}$
where $t\simeq -3$eV, $\left<\ve r,\ve r'\right>$ runs over nearest neighbors of a hexagonal lattice $\ve r = n_1\ve a_1+n_2\ve a_2+s\ve R_1$ ($n_1,n_2\in\mathbb Z, \sms s=0,1$) with $\ve a_1,\ve a_2$ the Bravais lattice vectors and $\ve R_1$ the vector connecting the A and B sublattices. $c_{\ve r}$ is a $\pi$-electron annihilation operator. We want to describe geometries which are lattice-translationally invariant along a zigzag direction, which we choose to be parallel to $\ve a_1$ (henceforth we call this direction the x-direction). Thus, it is convenient to transform to electron operators $d_{n, k,s} = N_x^{-\frac12} \sum_{n_1} e^{-i k n_1} c_{n_1,n,s}$, where $N_x$ is the number of unit cells in the ribbon along the x-direction, so that the 1D Brillouin zone is $[0,2\pi]$. In terms of the $d$-operators, the Hamiltonian reads
\begin{equation}
H_G = t \sum_{k,n}d^\dagger_{n,k,A}d_{n,k,B} + d^{\dagger}_{n,k,A}u_k d_{n-1,k,B} + h.c.,
\end{equation}
where $n$ labels the position along the $\ve a_2$ direction and $u_k = 1+e^{i k}$. A structural $\alpha$-edge is created by removing all terms with operators corresponding to positions $n<0$ and the terms with operators corresponding to the A sublattice site of $n=0$. For a $\beta$-edge only the $n<0$ terms need to be removed. A Hamiltonian truncated in such a way gives rise to exponentially localized, zero-energy edge states \cite{edge_states_theor}
\begin{equation}
\left|\psi_0^{\alpha/\beta}(k)\right> = \mathcal N_k^{\alpha/\beta} \sum_{n=0}^\infty e^{-n/\xi_k^{\alpha/\beta} + i n \phi} d^\dagger_{n,k,B/A} \left|0\right>
\end{equation}
with the normalization $\mathcal N_k^{\alpha,\beta} =  (1 - |u_k|^{\pm2})^{\frac12}$, the localization length $\xi_k^{\alpha,\beta} = \mp \left[\ln|u_k|\right]^{-1}$ and some unimportant phase $\phi=\arg(-u_k^{\pm1})$. Obviously, the $\alpha$-edge state only exists for $k\in\left[\frac{2\pi}3,\frac{4\pi}3\right]$ while the remaining k-space supports the $\beta$-edge state. Exactly at the projection of the valleys K and K' to the one-dimensional Brillouin zone, i.e. K$\rightarrow\frac{2\pi}3$ and K'$\rightarrow\frac{4\pi}3$, the localization lengths of both edge states diverge as $\frac2{\sqrt3}|k-{\rm K}|^{-1}$ or $\frac2{\sqrt3}|k-{\rm K'}|^{-1}$, respectively. For a ribbon with $\alpha$- and $\beta$-edges on opposite sides, a zero energy mode is present throughout the whole Brillouin zone. The corresponding wave function, however, `jumps' from one edge to the other as K or K' is crossed.

Now, instead of a structural graphene edge we want to describe a GG interface (we discuss only the $\alpha$-interface, the $\beta$-interface being analogous, see Appendix A). The essential features are well described within a simplified model which includes only one -C-H group at each `tooth' of the zigzag edge, instead of the whole graphane lattice. We further neglect all $\sigma$-orbitals of the carbon atoms. The Hamiltonian of this -C-H group reads
\begin{equation}
H_I= t d^\dagger_{k,C} d_{0,k,B} + t' d^\dagger_{k,H} d_{k,C} + h.c. +\epsilon_H d^\dagger_{k,H} d_{k,H}
\end{equation}
and must be added to the $\alpha$-truncated $H_G$, where $d_{k,C}$ and $d_{k,H}$ are the annihilation operators of the carbon $\pi$-orbital and the hydrogen s-orbital, respectively. Typically, $\epsilon_H\simeq -0.4eV$ is much smaller than the other two energy scales $t\simeq -3eV$ and $t'\simeq -5.8eV$ (see Appendix B). Since we are only interested in how the -C-H group affects the edge state, we project $H_G+H_I$ onto the subspace $\{\left|\psi_0^\alpha(k)\right>,\left|C\right> = d_{k,C}^\dagger\left|0\right>, \left|H\right> = d_{k,H}^\dagger\left|0\right>\}$. The projected Hamiltonian reads
\begin{equation}
H_{\rm proj.} = \left(\begin{matrix} 0 & t \mathcal N^\alpha_k & 0\\ t\mathcal N^\alpha_k & 0 & t'\\0 & t' & \epsilon_H\end{matrix}\right).
\end{equation}
In leading order perturbation theory in $\frac{\epsilon_H}{t'}$ and $\frac{t}{t'}$, this projection yields a low-energy edge state $\left|\psi^\alpha(k)\right> \sim t'\left|\psi_0^\alpha(k)\right> - t\mathcal N^\alpha_k\left|H\right>+\mathcal O(\epsilon_H)\left|C\right>$ with energy
\begin{equation}
\epsilon^\alpha(k) \simeq \epsilon_H\frac{t^2}{ t'^2}(2\cos(k-\pi)-1),\bs k\in\left[\frac{2\pi}3,\frac{4\pi}3\right] \label{simplified_energy},
\end{equation}
where we have dropped terms of order $\frac{\epsilon_H^2}{t'^2}$ and $\frac{t^4}{t'^4}$. Note that the amplitude of the state $\left|\psi^\alpha(k)\right>$ is of order $\epsilon_H$ at the C atom of the -C-H group. As a result, the inclusion of more and more graphane rows, i.e. -C-H groups, only leads to terms in the energy which are higher order in $\epsilon_H$. Thus, this simplified model is expected to describe the energy dispersion correctly as long as no other states which we have neglected here come too close in energy. Eq. (\ref{simplified_energy}) roughly describes a parabola around $k=\pi$ which crosses zero at K and K'. Therefore, the edge state at a GG interface has - in contrast to the edge state at a structural zigzag edge - a finite velocity, the magnitude of which is largely determined by the energy of the hydrogen 1s-orbital $\epsilon_H$ relative to the $\pi$-orbital energy in graphene (set to zero here). Note that also other effects which are not included in our simplified model, e.g. next-nearest neighbor hoppings \cite{nnn_dispersion} or local electrostatic gates at the edges \cite{gate_dispersion}, can increase the bandwidth of the edge state. For typical tight-binding parameters we find edge state velocities near K or K' of $10^4 - 10^5 \frac{\rm m}{\rm s}$ (see Appendix A).

The bandwidth of the edge state is crucial for the question of magnetic ordering at the edges. While conventional graphene edges are believed to be spin-polarized in their ground state \cite{ref_spin_polarization,ref_spin_polarization2}, the edge states considered here have a larger bandwidth which helps to suppress the magnetic phase transition. The bandwidth can be tuned further by gates \cite{gate_dispersion,fut_publication}.

\begin{figure}[!h]
\centering
\includegraphics[width=250pt]{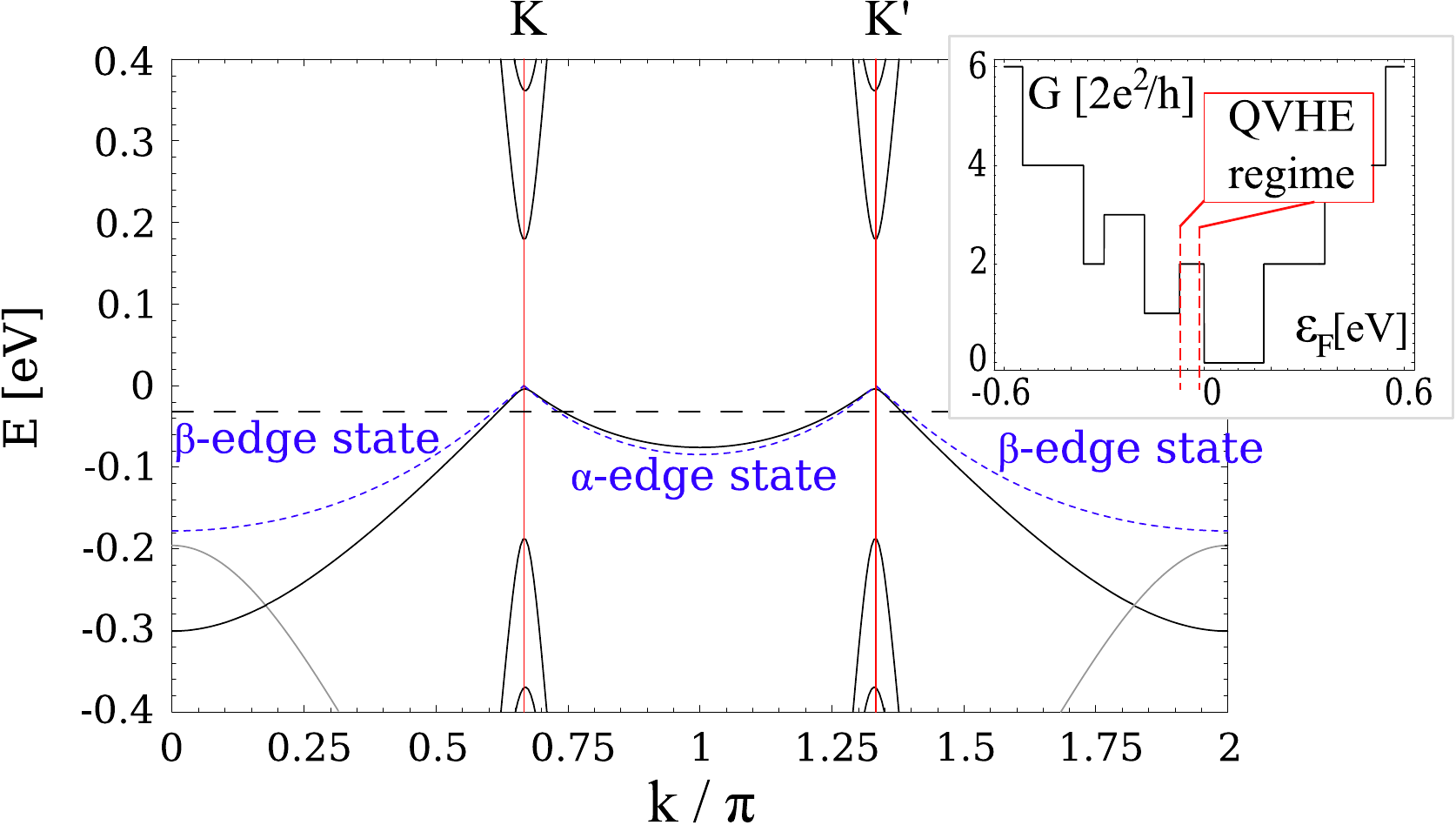}
\caption{(color-online) The band structure of a graphane-terminated $\alpha\beta$-ribbon with $W=10$nm. The dashed horizontal line is a typical Fermi energy $\epsilon_F$ for the QVHE regime chosen such that $\xi_{k_F}\ll W/2$. The gray bands correspond to exponentially localized states at the outer graphane edge which are not important here. The full curves are results of numerical calculations within the extended tight-binding model. The dashed (blue) curves show the effective model dispersion (\ref{simplified_energy}). The inset shows the conductance $G$ of the graphene region as a function of $\epsilon_F$.}
\label{fig_state}
\end{figure}

Fig. \ref{fig_state} compares the energy dispersion of the edge states calculated from the extended tight-binding model, as introduced below, with Eq. (\ref{simplified_energy}) and its analog for the $\beta$-edge state (see Appendix A). While the simplified model for the $\alpha$-edge state agrees well with the extended model, the simplified $\beta$-edge state dispersion predicts a too small bandwidth. However, the main features, namely the different directions of motion of electrons in the valley K (K') at different interfaces, are described properly by the simplified model.

{\it Edge state transport.} A ribbon with combined $\alpha$- and $\beta$-interfaces ($\alpha\beta$-ribbon) with finite width $W$ of the graphene region gives rise to a quantum valley Hall effect (QVHE) \cite{qvhe_footnote}, in which K (K') valley electrons move left (right) at the $\alpha$-interface and right (left) at the $\beta$-interface, if the Fermi energy $\epsilon_F$ is tuned as indicated in Fig. \ref{fig_state}. For wide enough ribbons ($\xi_{k_F}\ll  W/2$), the overlap between edge states moving in different directions is small so that backscattering is forbidden if valley scattering can be neglected. Since valley scattering requires the transfer of a large crystal momentum, only atomic scale disorder would destroy the QVHE.

The detection of this QVHE by a transport measurement could be a first experimental step to reveal the GG interface quality. In a clean $\alpha\beta$-ribbon the conductance is $\frac{2e^2}h n$, where $n$ is the number of transport modes intersected by $\epsilon_F$ (see inset of Fig. \ref{fig_state}).

For wide ($W\gg50$nm; see Appendix A) $\alpha\beta$-ribbons at temperatures below the spin-orbit gap in bulk graphene ($\sim10$mK \cite{spin_orbit_macdonald}) a new quantized Hall effect arises: the QSVHE. Here, not only the spin or the valley determines the direction of motion at the edges, like in the QSHE \cite{kane_qshe} or in the QVHE, respectively, but both, {\it spin and valley} are responsible for the selection of the edge and the direction of motion, if $\epsilon_F$ is tuned into the bulk SOI-induced gap in graphene. For instance, at an $\alpha$-interface only electrons in valley K and with spin pointing into a certain direction (we define this direction as $\uparrow$) are allowed to move left; the electron in valley K' and with spin pointing in the opposite direction ($\downarrow$) moves right. The other two combinations of valley and spin are not allowed. This increases the stability of the QSVHE compared to the pure QSHE or QVHE because backscattering requires simultaneous valley and spin scattering. In principle the QSVHE already allows spin-valley conversion, but only at very low temperatures ($\sim$10mK).

{\it Enhanced spin-orbit interaction.} 
It is well known that the smallness of the SOI in graphene is rooted in the lightness of carbon on one hand and in the symmetry-induced decoupling of the $\pi$- and $\sigma$-bands, on the other. The latter issue can be overcome, however, by twisting the lattice. While the carbon atoms of the A and B sublattices are coplanar in graphene, they are `pushed' out of the plane in graphane due to the rehybridization sp$_2\rightarrow$ sp$_3$ which is caused by the presence of the additional hydrogen atoms \cite{graphane_theor_1}. Thus, the $\pi$-band is locally coupled to the $\sigma$-band at the interface. We therefore expect an enhanced SOI near GG interfaces.

In order to substantiate this expectation, we use an extended tight-binding model of GG heterostructures which respects the $\pi$-band as well as the $\sigma$-band of the carbon lattice (see Appendices). It is based on an environment-dependent tight-binding model for general hydrocarbons \cite{edtb_hydrocarbon} and takes the atomic carbon SOI into account \cite{spin_orbit_macdonald}. We find that the GG edge states exhibit a much larger spin splitting than conventional edge states (see Fig. \ref{fig_spin_splitting}). This can be understood for the $\alpha$-edge state (for the $\beta$-edge state, see Appendix F) by projecting the on-site SOI Hamiltonian
\begin{equation}
H_{SO} = i\Delta \sum_{\ve r} \sum_{\mu\nu\rho\tau\tau'} \epsilon^{\mu\nu\rho} c^\dagger_{\ve r,p_\mu,\tau} \sigma^\nu_{\tau\tau'} c_{\ve r,p_{\rho},\tau'},
\end{equation}
where $c_{\ve r,p_\mu,\tau}$ annihilates an electron at site $\ve r$ in orbital $p_\mu$ with spin $\tau$, $\sigma^{\mu}$ are the Pauli matrices for the electron spin, $\mu,\nu,\rho=x,y,z$, and $\epsilon^{\mu\nu\rho}$ is the Levi-Civita tensor, onto the two-dimensional spin-degenerate subspace of the $\alpha$-edge state $\left|\psi^\alpha(k);\tau\right>$ obtained by diagonalizing the extended tight-binding Hamiltonian without SOI. In a conventional edge state $\left|\psi_0^\alpha(k);\tau\right>$ only the $\pi$-orbitals are occupied so that $\left<\psi_0^\alpha(k);\tau|H_{SO}|\psi_0^\alpha(k);\tau'\right>\equiv 0$ and the SOI becomes a higher order effect \cite{spin_orbit_macdonald}. At a GG interface, however, the edge state acquires contributions from the $\sigma$-orbitals because of the out-of-plane twisting of the carbon atoms which leads to a mixing of the $\pi$- and the $\sigma$-band. As a result, the SOI becomes a first order effect.

\begin{figure}[!h]
\centering
\includegraphics[width=220pt]{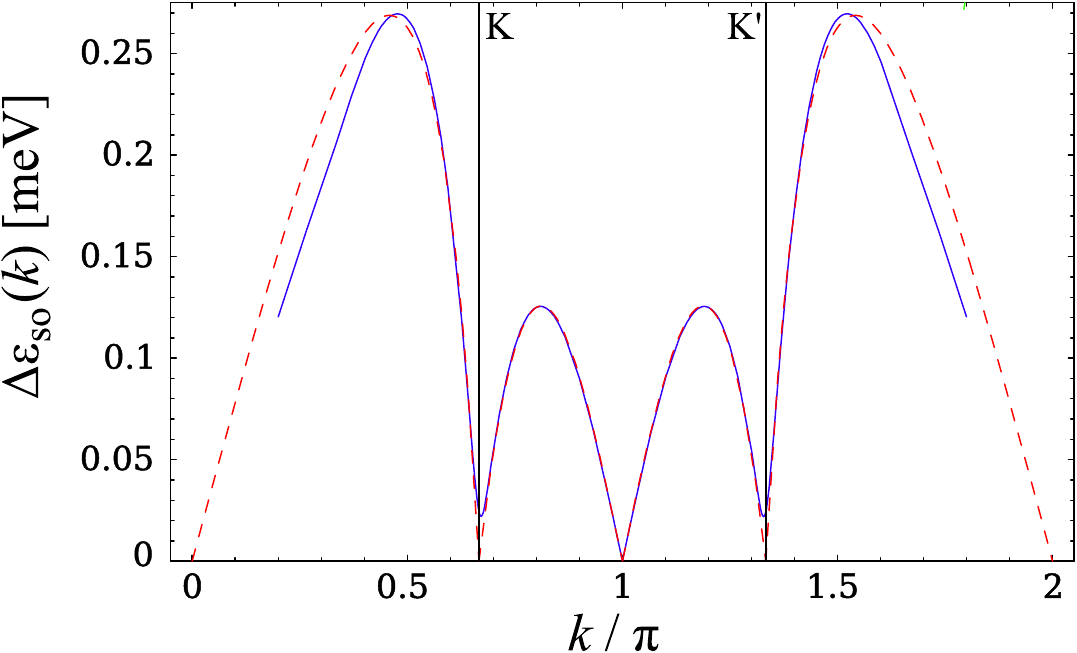}
\caption{(color-online) SOI induced spin splitting of the edge states. The solid (blue) curve shows the absolute value of the spin splitting $\Delta\epsilon_{SO}(k)$ of the edge states calculated from the extended tight-binding model which includes the bulk graphene SOI as well as the interface SOI. The dashed (red) line shows the spin splitting calculated from the approximate Hamiltonian (\ref{eff_so_ham}). The solid curve has been calculated for a 10nm wide $\alpha\beta$-ribbon. The spin splitting at K,K' which is larger than the spin-orbit splitting in bulk graphene is due to finite-size effects.}
\label{fig_spin_splitting}
\end{figure}

The projected Hamiltonian reads
\begin{equation}
H_{SO}^{\rm eff,\alpha} = \sum_{\tau\tau'}\int_{\frac{2\pi}3}^{\frac{4\pi}3} \frac{\D k}{2\pi} e_{k,\alpha,\tau}^\dagger \Gamma_{\tau\tau'}(k) e_{k,\alpha,\tau'},
\end{equation}
with $e_{k,\alpha,\tau}$ the $\alpha$-edge state annihilation operators and
\begin{eqnarray}
\Gamma_{\tau\tau'}(k) &=& \left<\psi^\alpha(k);\tau|H_{SO}|\psi^\alpha(k);\tau'\right> \\
&\simeq & (k-\pi) (\mathcal N_k^\alpha)^2 \left[\Delta_R^\alpha \sigma^y + \Delta_i^\alpha \sigma^z \right]_{\tau\tau'} \label{eff_so_ham},
\end{eqnarray}
where $\Delta_R,\Delta_i$ are constants describing the Rashba and intrinsic parts of the effective SOI. The Dresselhaus term ($\propto\sigma^x$) vanishes because of the mirror symmetry $x\rightarrow -x$ of the interfaces. The additional factor $(\mathcal N^\alpha_k)^2$ accounts for the fact that the effective SOI, generated by a GG interface, must be proportional to the amplitude of the state at the interface. The parameters $\Delta_R^{\alpha} = -0.16$meV and $\Delta_i^\alpha = -0.05$meV are obtained from a fit to the numerical results (see Appendix F). Because of time-reversal invariance the spin splitting is exactly zero at $k=0,\pm\pi$. These are also the points where the spin direction of the energetically higher edge state changes abruptly. Note that for wide ribbons ($W \gtrsim 100$nm), the spin splitting at K,K' is essentially given by the bulk SOI because $\mathcal N_k^{\alpha/\beta}$ vanishes for $k\rightarrow$K,K'. As a result, the SOI enhancement at the interfaces does not increase the critical temperature at which the QSVHE can be observed.

\begin{figure}[!h]
\centering
\includegraphics[width=220pt]{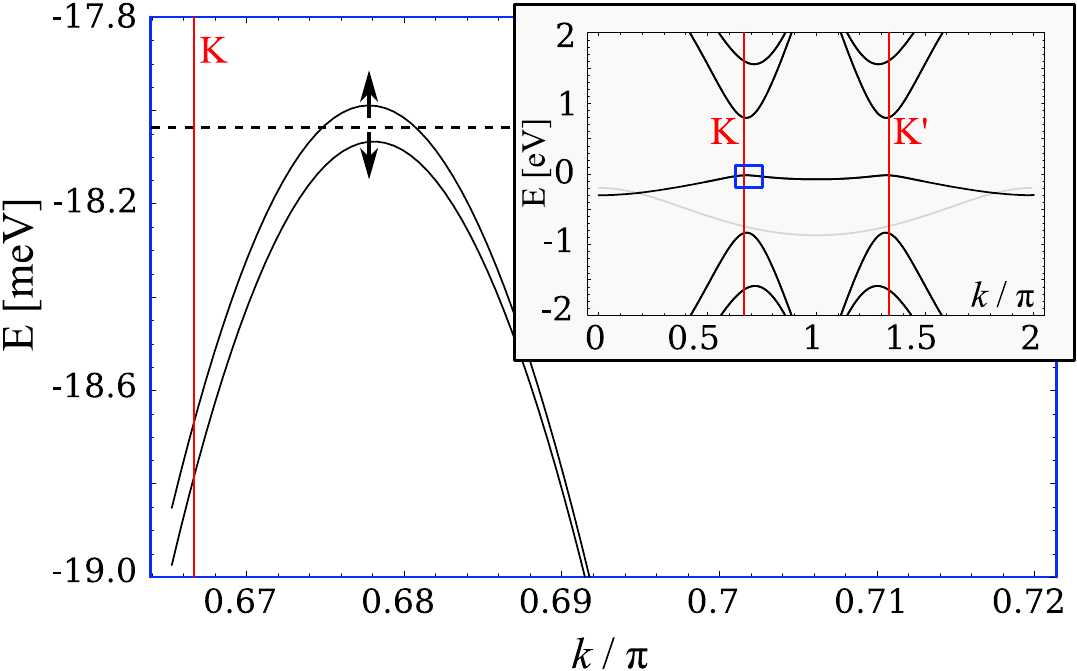}
\caption{Energy dispersion of the edge states of a 2nm wide $\alpha\beta$- ribbon near the K-valley. The two bands have opposite spin direction. The dashed line indicates a typical $\epsilon_F$. The inset shows a larger region of the band structure. Grey lines see caption of Fig. \ref{fig_state}.}
\label{fig_bandstructure}
\end{figure}

However, there is another possibility to use the interfacial enhancement of the SOI for high temperature spin-valley conversion. The spin splitting at K and K' decays like $W^{-1}$. 
Thus, one should use narrow ribbons in order to exploit the SOI enhancement at GG interfaces. In Fig. \ref{fig_bandstructure}, the band structure of a narrow GG heterostructure is shown. At the maxima of the edge mode the spin-orbit splitting is about 78 $\mu$eV $\simeq 0.9$K. Note that although these states are derived from edge states, they do not actually appear like edge states in this case because $\xi_k\gg W$ ($\xi_k$ diverges near K and K'). Thus, by squeezing the wide edge states into a narrow ribbon, the spin-orbit splitting is enhanced at K and K'. 

As required, the energetically higher band at K has the opposite spin-direction than the corresponding band at K'. Thus, by tuning $\epsilon_F$ as indicated in Fig. \ref{fig_bandstructure}, a spin-valley filter is realized: left- and right-movers exist for each spin and valley but the transmission of this structure is only non-zero for (K,$\uparrow$) or for (K',$\downarrow$) and zero for the other two combinations of valley and spin. This means that each spin-up electron which initially consists of K and K' components (valley-unpolarized) will be in a pure valley state (K here) after it has passed the narrow $\alpha\beta$-ribbon in x-direction, while a valley-unpolarized spin-down electron will be in a pure K' state after the passage. 

In conclusion, we have demonstrated that graphane can be utilized to create an effective edge for the $\pi$-band in graphene and that the SOI of the resulting edge states is strongly enhanced, compared to pure graphene. 
Furthermore, we have shown that, due to enhanced SOI at graphene/graphane interfaces, binary information which is encoded in the spin of an electron can be transferred to a valley-based encoding.

We acknowledge useful discussions with B. Braunecker, I. Martin, B. Trauzettel, and A. Yacoby. This work has been supported by the Swiss NF and the NCCR Nanoscience Basel.

\appendix

\section{Simplified model for the edge state}
\begin{figure}[!h]
\centering
\includegraphics[width=230pt]{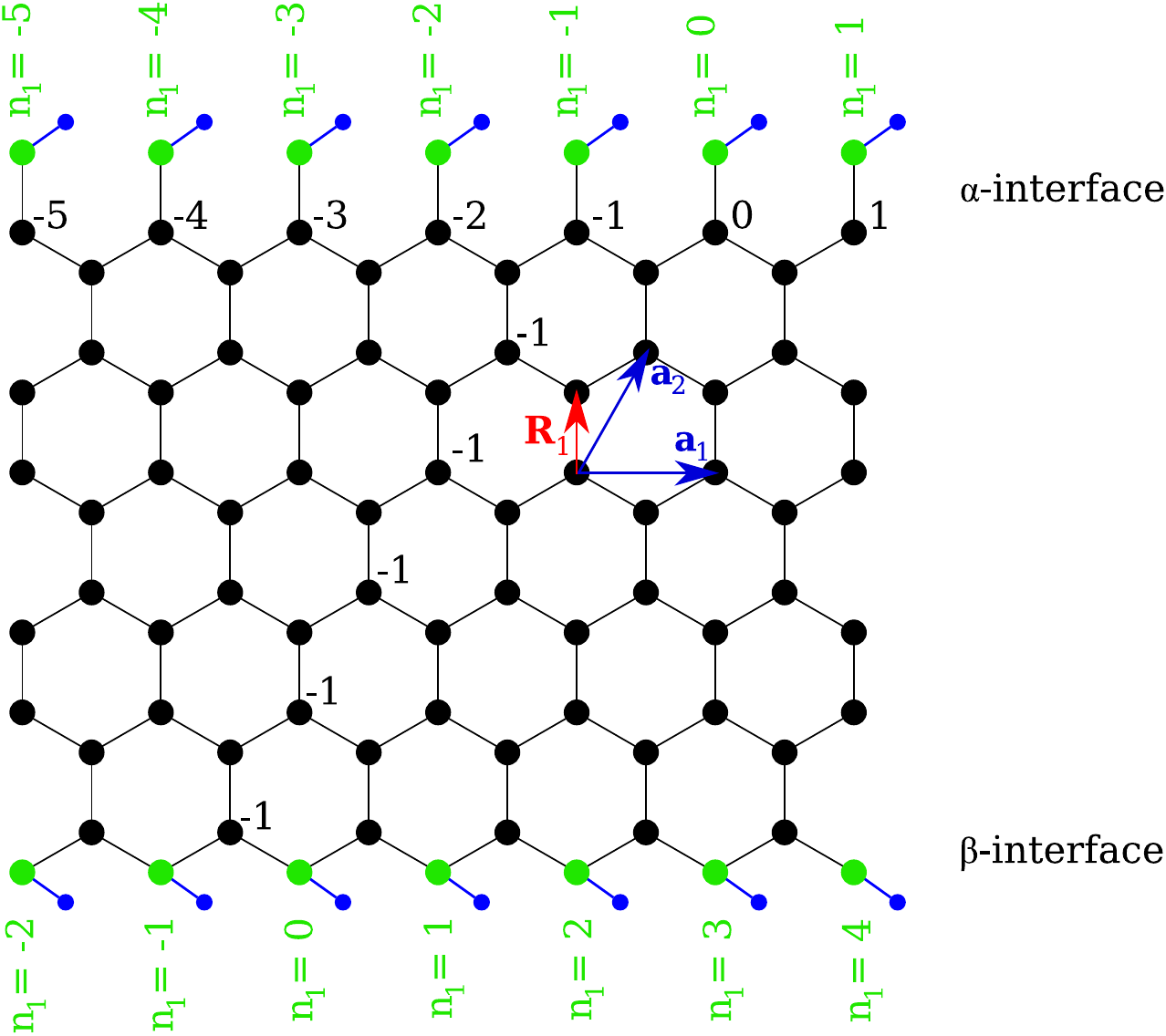}
\caption{{\bf Definitions in the simplified interface model.} The green dots represent the carbon atoms of the -C-H groups and the blue dots represent the hydrogen atoms. The black numbers indicate the $n_1$ coordinate of the carbon atoms in the graphene region.}
\label{supp1}
\end{figure}

The $\alpha$- and $\beta$-truncated Hamiltonians for the structural graphene zigzag edges read
\begin{eqnarray}
H_G^\alpha &=& t \sum_{k} \sum_{n=1}^\infty \left[ d_{n,k,A}^\dagger d_{n,k,B} + d_{n,k,A}^\dagger u_k d_{n-1,k,B} \right] \\
H_G^\beta &=& t \sum_{k} \left\{ \sum_{n=1}^\infty \left[  d_{n,k,A}^\dagger d_{n,k,B} + d_{n,k,A}^\dagger u_k d_{n-1,k,B} \right] \right.\nonumber\\
&&\left. + d_{n,k,A}^\dagger d_{n,k,B} \right\} ,
\end{eqnarray}
with $u_k=1+e^{ik}$ and the simplified model Hamiltonians for the $\alpha$- and $\beta$-interfaces between graphene and graphane read
\begin{equation}
H_{I}^\alpha =  t d^\dagger_{k,C} d_{0,k,B} + t' d_{k,H}^\dagger d_{k,C} + h.c.  +\epsilon_H d_{k,H}^\dagger d_{k,H}
\end{equation}
\begin{equation}
H_I^\beta =  t u_k d^\dagger_{k,C} d_{0,k,B} + t' d_{k,H}^\dagger d_{k,C} + h.c. +\epsilon_H d_{k,H}^\dagger d_{k,H},
\end{equation}
where we have defined
\begin{eqnarray}
d_{k,H} &=& N^{-\frac12} \sum_{n_1} e^{-i k n_1} c_{n_1,H} \\
d_{k,C} &=& N^{-\frac12} \sum_{n_1} e^{-i k n_1} c_{n_1,C}.
\end{eqnarray}
The operators $c_{n_1,C}$ annihilate an electron in the $\pi$-orbital of the carbon atom next to the edge, as shown in Fig. \ref{supp1} and the operators $c_{n_1,H}$ annihilate an electron in the 1s-orbital of the corresponding hydrogen atom.

The projection of $H^{\alpha/\beta}_G + H^{\alpha/\beta}_I$ onto the subspaces $\left\{\left|\psi_0^{\alpha/\beta}(k)\right>,\sms \left|1\right> = d^\dagger_{k,C}\left|0\right>,\sms\left|2\right> = d^\dagger_{k,H}\left|0\right>\right\}$ reads in matrix form
\begin{eqnarray}
\left[H_G^\alpha + H_I^\alpha\right]_{\rm proj.} &=& \left[\begin{matrix}
0 & t \mathcal N^\alpha_k & 0 \\
t \mathcal N^\alpha_k & 0 & t' \\
0 & t' & \epsilon_H
\end{matrix}\right] \\
\left[H_G^\beta + H_I^\beta\right]_{\rm proj.} &=& \left[\begin{matrix}
0 & t u_k \mathcal N^\beta_k & 0 \\
t u_k^*\mathcal N^\beta_k & 0 & t' \\
0 & t' & \epsilon_H
\end{matrix}\right].
\end{eqnarray}
Because $|\epsilon_H|\ll |t|,|t'|$, we are only interested in results to order $\epsilon_H$. We find a low energy edge state for both $\alpha$- and $\beta$-interfaces
\begin{eqnarray}
\epsilon^\alpha &=& \frac{t^2 (\mathcal N^\alpha_k)^2}{t^2 (\mathcal N^\alpha_k)^2 + t'^2} \epsilon_H + \mathcal O(\epsilon_H^2),\sms k\in\left[\frac{2\pi}3,\frac{4\pi}3\right]\\
\epsilon^\beta &=& \frac{t^2 |u_k|^2 (\mathcal N^\beta_k)^2}{t^2 |u_k|^2(\mathcal N^\beta_k)^2 + t'^2} \epsilon_H + \mathcal O(\epsilon_H^2),\sms k\in\left[-\frac{2\pi}3,\frac{2\pi}3\right] \nonumber. \\
\end{eqnarray}
Near the boundary of the k-space domains of the edge states, K$=\frac{2\pi}3$ and K'$=\frac{4\pi}3$, the dispersion is approximately linear and we can write for both states ($\alpha/\beta$) in both valleys (K,K')
\begin{equation}
\epsilon^{\alpha/\beta} = \sqrt3\frac{t^2\epsilon_H}{t'^2} |\delta k| + \mathcal O(\delta k^2),
\end{equation}
where $\delta k = k - $K or $\delta k = k-$K'. From this we can estimate the typical velocity of the edge states near K and K' to
\begin{equation}
|v^{\alpha/\beta}| \simeq \sqrt 3 \frac{t^2 |\epsilon_H|}{t'^2} \frac{a_0 \sqrt 3}\hbar \simeq 6.8\cdot 10^4 \frac{\rm m}{\rm s},
\end{equation}
which is about one order of magnitude smaller than the Fermi velocity in graphene.

Next, we would like to estimate the typical localization length of the edge states in the QSVHE regime. If we take the SOI into account, a spin gap $\Delta\simeq 2\mu$eV opens up at K and K' for wide ribbons ($W\gtrsim100$nm). This spin gap is of the order of the spin-orbit splitting in bulk graphene because the edge states become completely delocalized over the graphene region at K and K'. If the Fermi energy is of the order of the spin gap, the four Fermi momenta can be estimated by
\begin{equation}
k_F^{1/2} \simeq {\rm K} \pm \Delta \frac{t'^2}{\sqrt3 t^2 \epsilon_H} \simeq {\rm K} \pm 3.4\cdot 10^{-3}\pi.
\end{equation}
\begin{equation}
k_F^{3/4} \simeq {\rm K'} \pm \Delta \frac{t'^2}{\sqrt3 t^2 \epsilon_H} \simeq {\rm K'} \pm 3.4\cdot 10^{-3}\pi.
\end{equation}
From this, the corresponding localization lengths for all interface states in the QSVHE regime are
\begin{equation}
\xi^{\alpha/\beta}_{k_F} = \frac32 a_0 |\ln|u_{k_F}||^{-1} \simeq 22{\rm nm},
\end{equation}
with $a_0\simeq 1.4$\AA\, the nearest neighbor C-C distance.

\section{Tight binding parameters}
We want to model the graphene/graphane heterostructures by a nearest-neighbor tight-binding model which takes into account the $\pi$-band and the $\sigma$-band of the hexagonal carbon backbone and the 1s-orbitals of the hydrogen atoms attached to each C atom in the graphane region. We neglect the non-orthogonalities of orbitals on different sites. First of all, we need the bare tight-binding hopping integrals between the oriented carbon orbitals 2s, 2p$_x$, 2p$_y$, 2p$_z$, i.e. the matrix elements of the Hamiltonian $H$
\begin{eqnarray}
V_{ss} &=& \left<2s;\ve r_0|H|2s; \ve r_1\right> \label{hop1}\\
V_{sp} &=& \left<2p_z;\ve r_0|H|2s;\ve r_1\right> \\
V_{pp}^{\sigma} &=& - \left<2p_z;\ve r_0|H|2p_z;\ve r_1\right> \\
V_{pp}^\pi &=& \left<2p_x;\ve r_0|H|2p_x;\ve r_1\right>
\end{eqnarray}
and the hopping integrals between carbon orbitals and the hydrogen 1s-orbital
\begin{eqnarray}
W_{ss} &=&\left<1s;\ve r_0|H|2s;\ve r_2\right> \\
W_{sp} &=&\left<2p_z;\ve r_0|H|1s;\ve r_2\right>\label{hopl},
\end{eqnarray}
where $\ve r_0 = (0,0,0)^T$, $\ve r_1=a_0(0,0,1)^T$ and $\ve r_2 = b_0(0,0,1)^T$ with $a_0$ the nearest-neighbor C-C distance and $b_0$ the C-H distance. The kets $\left|2s;\ve r\right>,\left|2p_i;\ve r\right>$ represent the 2s- and the 2p$_i$-orbitals ($i=x,y,z$) of the carbon atoms at $\ve r$. The ket $\left|1s;\ve r\right>$ represents the hydrogen 1s-orbital.

The H-H distance in the chair conformation of graphane is large enough to neglect the direct hopping between hydrogen orbitals. All nearest-neighbor hopping integrals between the relevant orbitals in graphane can be reduced to the parameters defined in Eqs. (\ref{hop1}-\ref{hopl}), as we will show subsequently.

In general, these bare hopping parameters are environment-dependent \cite{_edtb_carbon,edtb_hydrocarbon}. However, there is a considerable variance of hopping parameters in the literature. For instance, the very often used parameters in Ref. \onlinecite{saito_book} deviate considerably from the ones extracted from Ref. \onlinecite{_edtb_carbon} (see Tab. \ref{tab_parameters}). In this case, the deviation is probably due to the different scopes of Refs. \onlinecite{_edtb_carbon} and \onlinecite{saito_book}, namely the crystal structure and the electronic properties, respectively. Therefore, we believe that the carbon orbital hopping parameters of Ref. \onlinecite{saito_book} are more suitable to our needs. Furthermore, as Reich {\it et al.} argue \cite{reich} for the $\pi$-band, for a quantitative description one would have to include up to 3rd neighbor hoppings and the corresponding non-orthogonalities. These parameters are, however, to our best knowledge, not available for the $\sigma$-band in graphene. Since we only aim at a description on a qualitative level, we refrain from more quantitative first principles calculation. We rather check that our results are robust with respect to variations in the parameters.

\begin{table}[!h]
\begin{center}
\begin{tabular}{|c|c|c|} \hline
 & Ref. \onlinecite{_edtb_carbon} & Ref. \onlinecite{saito_book} (used here) \\ \hline \hline
$V_{ss}$ & -4.6 eV & -6.8 eV \\ \hline
$V_{pp}^\pi$ & -2.4 eV & -3.0eV \\ \hline
$V_{pp}^\sigma$ & -7.7eV & -5.0eV \\ \hline
$V_{sp}$ & -5.7eV & -5.6eV\\ \hline
\end{tabular}
\end{center}
\caption{Two sets of nearest-neighbor tight-binding parameters from the literature. We finally use the second column of parameters in the calculation.}
\label{tab_parameters}
\end{table}

Furthermore, we make the simplifying assumption that the bare hopping matrix elements are the same in the graphene and in the graphane region. This does not mean, of course, that the final Hamiltonian in the carbon subspace is equal for graphene and graphane. The different relative alignment of the A- and the B-sublattice is taken into account later.

The parameters for the C-H bond in the graphane region ($W_{ss}$ and $W_{sp}$) are calculated along the lines of Ref. \onlinecite{edtb_hydrocarbon}, with a C-H bond length of 1.1\AA\, \cite{sofo}. They are
\begin{equation}
W_{ss} = -5.4 {\rm eV},\bs W_{sp} = -5.8{\rm eV}.\label{hoph}
\end{equation}

Finally, the 2p-orbital energy of carbon defines the zero of the energy. The relative energy of the carbon 2s-orbital is $\epsilon_C^s=-8.7$ eV \cite{saito_book}. The orbital energy of the hydrogen 1s-orbital can be roughly estimated from Refs. \onlinecite{_edtb_carbon,edtb_hydrocarbon}. We assume that the orbital energies of the carbon orbitals is elevated due to the nearby hydrogen atom about the same amount as due to the other carbon atoms. In this work we use $\epsilon_H^s=-0.4$eV for the hydrogen 1s-orbital. The results and conclusions do not depend critically on this value.

\section{Orbital ribbon Hamiltonian}
We first want to derive the Hamiltonian, describing bulk graphane and bulk graphene, in order to check the resulting band structure against {\it ab initio} band structures in the literature \cite{sofo}. For doing this, we need to translate the bare hopping integrals (\ref{hop1}-\ref{hoph}) to the hoppings on the lattice under consideration. We define the honeycomb lattice vectors $\ve a_1 = a_0\left(\sqrt 3,0,0\right)^T$ and $\ve a_2 = a_0\left(\frac{\sqrt3}2,\frac32,0\right)^T$ and the nearest neighbor vectors
\begin{eqnarray}
\ve R_1 &=& a_0(0,1,z)^T \\ 
\ve R_2 &=& a_0\left(-\frac{\sqrt3}2,-\frac12,z\right)^T \\
\ve R_3 &=& a_0\left(\frac{\sqrt3}2,-\frac12,z\right)^T,
\end{eqnarray}
where $a_0$ is the C-C distance of flat graphene and $z a_0$ is the separation of the A- and B-sublattice planes. We chose the parameter $z=z_0\simeq 0.42$ in this work. At each carbon site, we define the system of the 2s- and 2p-orbitals in the following way: the s-orbital is spherically symmetric, so that the atomic alignment is irrelevant. the p-orbitals can be characterized by a vector, pointing into the direction of the positive part of the orbital wave function. The wave functions of the second carbon shell are approximately
\begin{eqnarray}
\psi_{2s}(\ve r) &=& f_{2s}(|\ve r|) \\
\psi_{2p_i}(\ve r) &=& f_{2p}(|\ve r|)  r_i,\sms i=x,y,z \label{p_orbital_basis}.
\end{eqnarray}
Thus, we choose the alignment of the three 2p-orbitals according to the same coordinate system which defines $\ve a_1,\ve a_2$. The set of p-orbitals transforms as a vector. This is useful if we want to express a p-orbital pointing, say, into the (1,1,0) direction. The wave function of this orbital can be written in terms of the basis functions (\ref{p_orbital_basis})
\begin{equation}
\psi_{2p,(1,1,0)}(\ve r) = \frac1{\sqrt2} \left[\psi_{2p_x}(\ve r)+\psi_{2p_y}(\ve r)\right].
\end{equation}
The direction of an orbital is conveniently expressed by a vector $\ve p$.

We need to deal with three different types of C-C hoppings. There is the hopping from an s-orbital to another s-orbital (ss), the hopping from an s-orbital to a p-orbital (sp) and the hopping between p-orbitals (pp). For the subsequent discussion we introduce the bond vector $\ve r_b$ which connects the two atoms participating in the bond under consideration. For an sp-bond $\ve r_b$ points always from the atom, carrying the s-orbital, to the atom on which the p-orbital is located. For ss-bonds and pp-bonds the direction of $\ve r_b$ plays no role.

The ss-hopping does not depend on the direction of $\ve r_b$ but only on the bond length. The bond length, however, is assumed constant over the whole lattice. In a homogeneous structure (only graphene {\it or} graphane), this assumption is fulfilled perfectly. In a heterostructure like the one we aim to describe there is a minor difference in the bond lengths in the different regions. We neglect this difference in the bond length because we believe that the dominant difference between graphene and graphane is the different symmetry of the lattice.

\begin{figure}
\centering
\includegraphics[width=180pt]{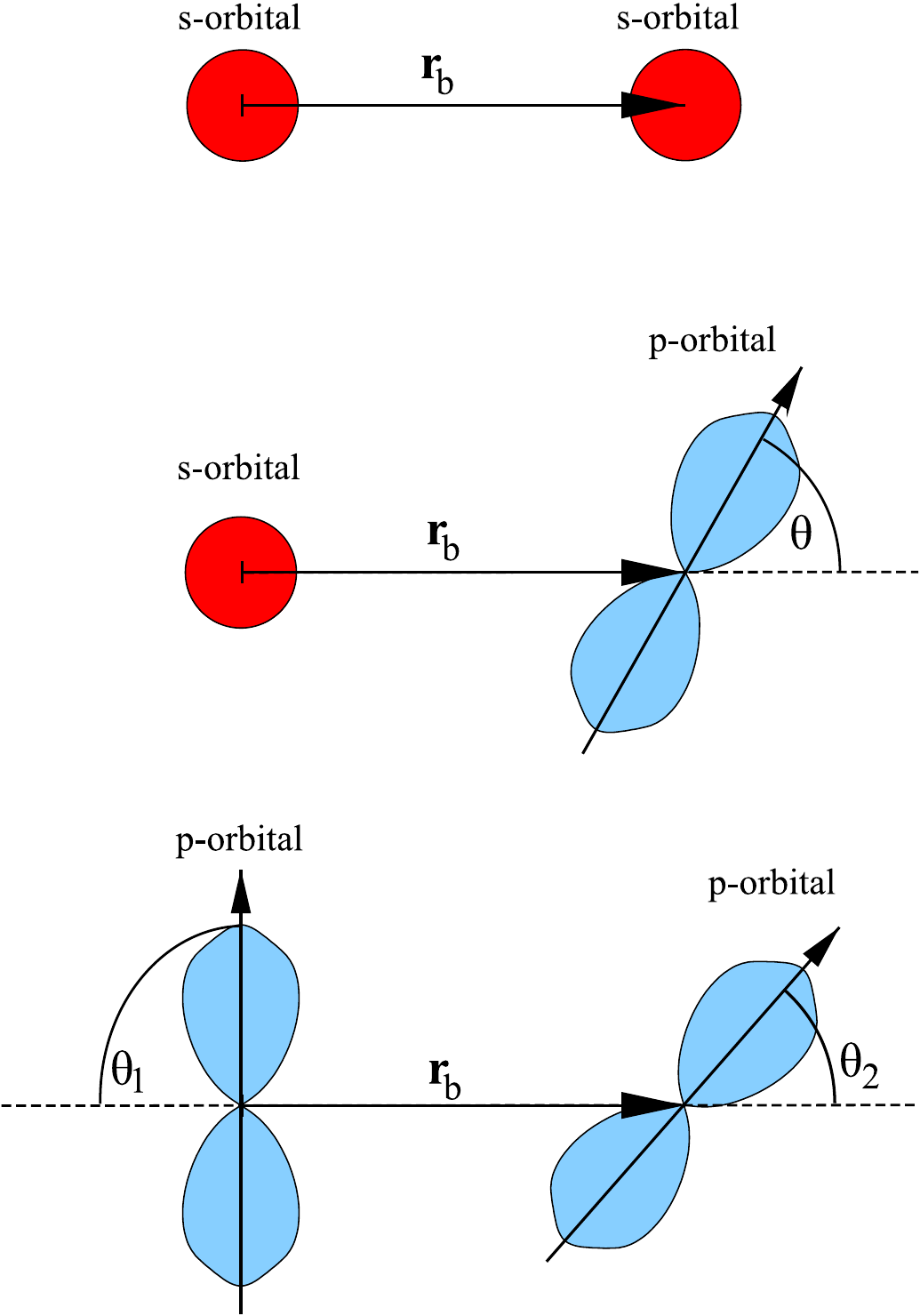}
\caption{Relative orientation of the carbon s- and p-orbitals. $\ve r_b$ is the bond vector, connecting the positions of the two atoms.}
\label{hopping_angles}
\end{figure}

For sp-hopping we only know the hopping integral if the p-orbital is aligned along the bond vector. The hopping between an s-orbital and a p-orbital that is perpendicular to the bond vector is zero by symmetry. Thus, only the angle $\theta$ between the p-orbital and the bond vector is important (see Fig. \ref{hopping_angles}). The sp-hopping integral of such a bond is then
\begin{equation}
t_{sp}(\theta) = -V_{sp}\cos\theta,\bs\cos\theta = \frac{\ve r_b\cdot \ve p}{|\ve r_b||\ve p|}. \label{sp_hopping}
\end{equation}

For pp-hopping, three angles are relevant: the two angles $\theta_1,\theta_2$ between the p-orbitals $\ve p_1,\ve p_2$ and the bond-vector $\ve r_b$ (see Fig. \ref{hopping_angles}) and the angle $\rho$ between the planes spanned by $\ve r_b,\,\ve p_1$ and $\ve r_b,\,\ve p_2$. The hopping integral of such a bond is
\begin{equation}
t_{pp} (\theta_1,\theta_2,\rho) = \cos\theta_1\cos\theta_2 V_{pp}^\sigma + \sin\theta_1\sin\theta_2\cos\rho V_{pp}^\pi \label{pp_hopping}
\end{equation}
with
\begin{equation}
\cos\theta_1 = -\frac{\ve r_b\cdot \ve p_1}{|\ve r_b||\ve p_1|},\bs \cos\theta_2 = \frac{\ve r_b\cdot \ve p_2}{|\ve r_b||\ve p_2|}
\end{equation}
and
\begin{equation}
\rho = \sphericalangle[\ve r_b\times \ve p_1 , \ve r_b\times\ve p_2].
\end{equation}

With the above equations we can evaluate all hopping matrix elements we need to write down the Hamiltonian. We start with the Hamiltonian in the carbon subspace $H_C$. We denote the sublattice by $s=A,B$, the Bravais lattice vector $\ve R_{\ve n} = n_1\ve a_1+n_2\ve a_2$ and the orbitals by $\mu=0,1,2,3$ for the s-, p$_x$-, p$_y$-, p$_z$-orbital, respectively. The real-space Hamiltonian reads
\begin{widetext}
\begin{equation}
H_C = \sum_{\ve n,\mu,\mu'} c^\dagger_{\ve n,\mu,A} \left[ t_{\mu,\mu',1} c_{\ve n,\mu',B} + t_{\mu,\mu',2} c_{\ve n - (0,1),\mu',B} + t_{\mu,\mu',3} c_{\ve n+(1,-1),\mu',B} \right] + h.c.\;,
\end{equation}
where $c_{\ve n,\mu,s}$ are electron annihilation operators and $t_{\mu,\mu',j}$ for $j=1,2,3$ are the hopping integrals between orbitals $\mu$ and $\mu'$ calculated by Eqs. (\ref{sp_hopping}),(\ref{pp_hopping}) for a bond between a central A atom and its three neighboring B atoms. For a bulk system, this Hamiltonian can be transformed to k-space as usual
\begin{equation}
H_C = \sum_{\ve k} \left\{\epsilon_s^C \left[d^\dagger_{\ve k,0,A} d_{\ve k,0,A} + d^\dagger_{\ve k,0,B} d_{\ve k,0,B} \right]+ \left[\sum_{\mu,\mu'} d^\dagger_{\ve k,\mu,A} f_{\mu\mu'}(\ve k) d_{\ve k,\mu',B} + h.c. \right]\right\}\label{ham_c_bulk},
\end{equation}
\end{widetext}
where
\begin{equation}
f_{\mu\mu'}(\ve k) = t_{\mu,\mu',1} + t_{\mu,\mu',2} e^{-i k_2} + t_{\mu,\mu',3}e^{i(k_1-k_2)}
\end{equation}
and
\begin{equation}
d_{\ve k,\mu,s} = \frac1{\sqrt N}\sum_{n_1,n_2} e^{-i(k_1n_1+k_2n_2)}c_{\ve n,\mu,s}.
\end{equation}
Note that $k_1,k_2$ are the components of the k-space coordinate w.r.t. the reciprocal vectors, which are non-orthogonal. In terms of $k_x$ and $k_y$ we have $k_1 = \sqrt 3 k_x$ and $k_2 = \frac12(\sqrt 3 k_x+3 k_y)$.

\begin{figure}
\centering
\includegraphics[width=220pt]{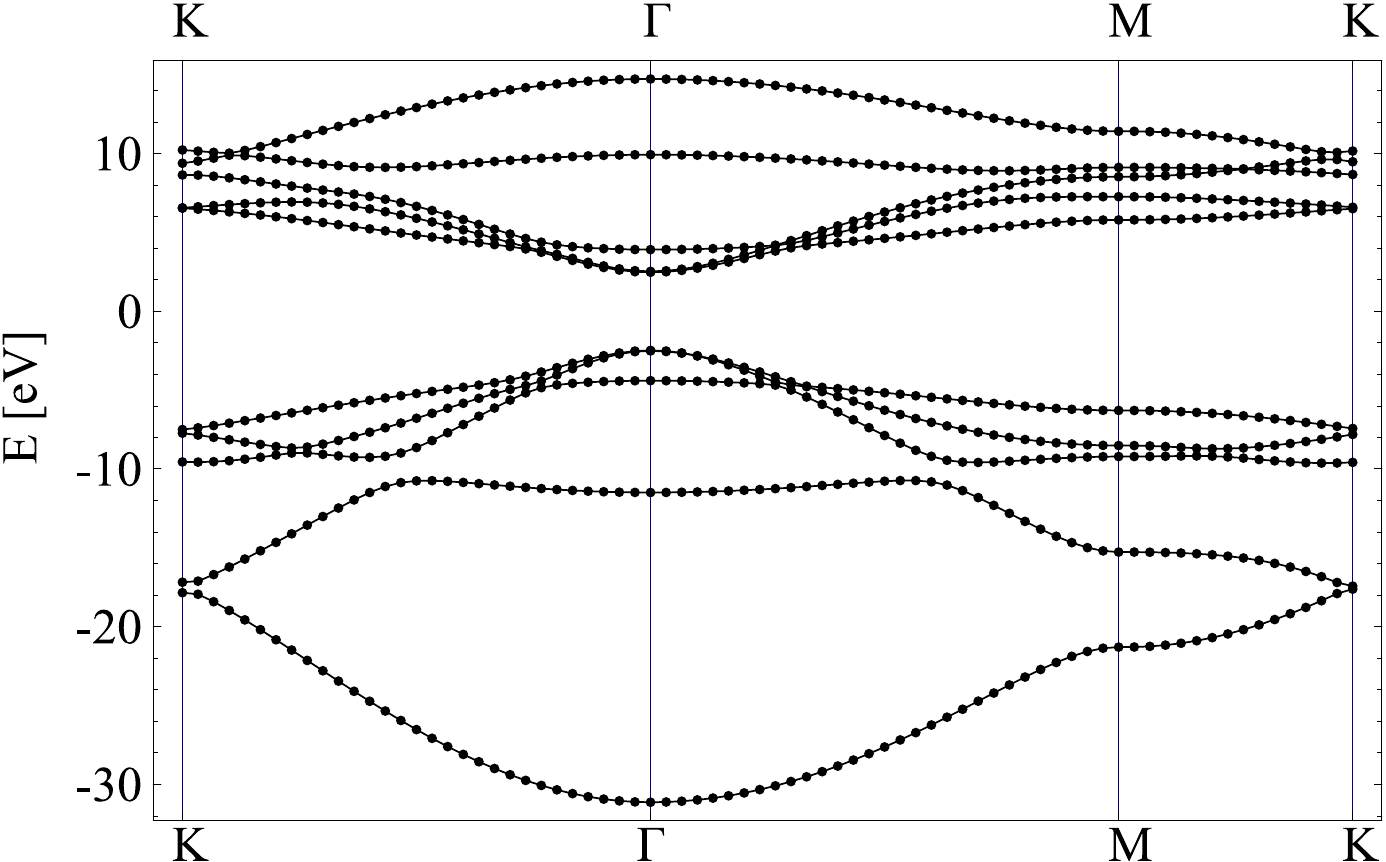}
\caption{Bulk band structure of graphane calculated from our tight binding model (Eqs. (\ref{ham_c_bulk}) and (\ref{ham_hc_bulk})).}
\label{graphane_band_structure}
\end{figure}

The graphene regions of the heterostructures are fully described by $H_C$. In the graphane region, the hydrogen-related terms must be added to $H_C$. Because we only consider nearest-neighbor hopping and the chair conformation of graphane, direct inter-hydrogen hopping is not allowed. Since the hydrogen atoms are supposed to sit nicely on top ($z>0$) of the B sites and below the A sites ($z<0$), the only non-zero hopping integrals, allowed by symmetry, are the two between the hydrogen 1s-orbital and the carbon 2s- and 2p$_{z}$-orbitals, i.e. $W_{ss}$ and $W_{sp}$, respectively. Because of this perfect alignment, no further transformation of the hopping integrals is needed. Only, because the 2p$_{z}$-orbital points in positive $z$-direction at the A- as well as at the B-sites, we have to respect the different sign of the sp-hopping between the carbon atoms on A-sites and their attached hydrogen atoms and those on B-sites. We find
\begin{widetext}
\begin{equation}
H_{H-C} = \sum_{\ve k,s} \left\{ \epsilon_s^C d^\dagger_{\ve k,H,s} d_{\ve k,H,s} + \left[W_{ss} d^\dagger_{\ve k,H,s} d_{\ve k,0,s} + (-1)^s W_{sp} d^\dagger_{\ve k,H,s} d_{\ve k,3,s}+h.c.\right]  \right\}.\label{ham_hc_bulk}
\end{equation}
\end{widetext}
The spinless bulk Hamiltonian $H = H_C + H_{H-C}$ can be represented as a $\ve k$-dependent $(10\times10)$-dimensional matrix which is easily diagonalized. Fig. \ref{graphane_band_structure} shows the graphane band structure calculated from $H$. Compared to the first-principles band structure of Sofo {\it et al.} \cite{sofo}, the band width of the lowest band is somewhat larger in our tight-binding model. This is typical for tight-binding models on honeycomb lattices with only nearest-neighbor hoppings and orthogonal orbital wave functions. The band gap is with $E_g\simeq 4.9 $eV also larger than the one found in Ref. \onlinecite{sofo}. However, Lebegue, {\it et al.} \cite{lebegue}, point out that the band gap is rather 5.4 eV. The band gap of 3.5 eV, as found by Sofo, is, according to Lebegue {\it et al.}, due to the use of the generalized gradient approximation which is known to predict erroneous energy gaps.

For modeling graphane-terminated graphene nanoribbons which are lattice-translationally invariant along the $\ve a_1$-direction (zigzag ribbons), we perform a partial Fourier transform of the carbon-electron operators, i.e.
\begin{equation}
d_{k,\mu,n,s} = \frac1{N_y} \sum_{n'} e^{-i k n'} c_{n',n,\mu,s}.
\end{equation}
The ribbon Hamiltonian in the carbon subspace then reads
\begin{widetext}
\begin{equation}
H_C = \sum_{k,n,\mu,\mu'} d^\dagger_{n,k,\mu,A} \left[t_{\mu,\mu',1} d_{n,k,\mu',B} + \left(t_{\mu,\mu',2} + e^{ik}t_{\mu,\mu',3}\right)d_{n-1,k,\mu',B} \right] + h.c.
\end{equation}
\end{widetext}
It is important to note that the hopping integrals $t_{\mu,\mu',j}$ are different in the graphane and graphene regions. At the interfaces we assume that the carbon atoms in the last graphane row have the sp$^3$-like orbital configuration of graphane rather than the sp$^2$ configuration of graphane. There is some ambiguity in this choice. Different choices of the interface properties do not lead to qualitatively different results. Only the magnitude of the interface-induced spin-orbit splitting may be renormalized by factors of order one.

The hydrogen part of the Hamiltonian is easily added to the graphane regions, as explained above.

\section{Spin-orbit Hamiltonian}
We restrict ourselves to the on-site spin-orbit interaction generated by the Hamiltonian
\begin{equation}
H_{SO}=\underbrace{\frac{\hbar}{4m^2 c^2}}_{\equiv A} (\nabla V(\ve r)\times \hat{\ve p})\cdot \boldsymbol \sigma,
\end{equation}
where $V(\ve r)$ is the (a priori unknown) potential of the carbon ions, $\hat{\ve p}=-i\hbar(\partial_x,\partial_y,\partial_z)$ is the momentum operator of an electron, and $\boldsymbol \sigma=(\sigma^x,\sigma^y,\sigma^z)$ are the Pauli matrices for the electron spin. We are interested in the matrix elements of $H_{SO}$ in the subspace spanned by the carbon 2s- and 2p-orbitals, i.e. the states $\left|\psi_{\mu,\tau}\right>\equiv\left|\mu,\tau\right>\equiv\left|\mu\right>\otimes\left|\tau\right>$ where $\mu=0,1,2,3$ labels the s-, p$_x$-, p$_y$-, p$_z$-orbital, respectively, and $\tau=\uparrow,\downarrow$ labels the spin.
\begin{equation}
\left<\mu,\tau|H_{SO}|\mu',\tau'\right> = A \sum_{i=x,y,z}\left<\mu|\left[\nabla V(\ve r)\times\hat {\ve p}\right]_i|\mu'\right>  \sigma^i_{\tau\tau'}.
\end{equation}
Now we perform an explicit calculation of the matrix elements and relate all non-zero integrals to each other by symmetry considerations. In real space, we can write ($i,j,k=x,y,z$)
\begin{equation}
\left<\mu|\left[\nabla V(\ve r)\times\hat {\ve p}\right]_i| \mu'\right> = -i\epsilon^{ijk} \int \D^3\ve r  \,\psi_\mu^*(\ve r) (\partial_j V) \partial_k \psi_{\mu'}(\ve r),
\end{equation}
where we used the wave functions defined in Eq. (\ref{p_orbital_basis}). From the symmetry of the orbital wave functions and the potential $V(\ve r)$ it is easy to see that
\begin{eqnarray}
\left<s|H_{SO}|p_i\right> = \left<p_i|H_{SO}|s\right> &=&0 \\
A\left<p_i\left|\left[\nabla V\times \hat {\ve p}\right]_j\right|p_k\right> &=&  i \epsilon^{ijk}\Delta,
\end{eqnarray}
where $\epsilon^{ijk}$ is the Levi-Civita tensor and
\begin{equation}
\Delta =\frac{\hbar}{4 m^2 c^2} \int_0^\infty \D r\, r |f_p(r)|^2 V'(r) .
\end{equation}
In second quantization the Hamiltonian reads
\begin{equation}
H_{SO} = i\Delta \sum_{\ve n,s}\sum_{\substack{\mu,\nu,\eta\\ \tau\tau'}}\epsilon^{\mu\nu\eta}c_{\ve n,\mu,s,\tau}^\dagger \sigma_{\tau\tau'}^\nu c_{\ve n,\eta,s,\tau'} \label{so_hamiltonian},
\end{equation}
where $\mu,\nu,\eta=x,y,z$. The coupling constant $\Delta\simeq 3meV$ is determined from the atomic spin-orbit interaction (see, e.g., Ref. \onlinecite{spin_orbit_macdonald}).

\section{Graphane-terminated $\alpha\beta$-ribbons}
As mentioned above, there is some ambiguity in the alignment of the interface bonds. There are two extreme cases. The first is shown in Fig. \ref{fig_alpha_beta_interface}: there, the bonds between the graphene and the graphane region are aligned as if the first graphene atom was sp$^3$-hybridized, i.e. they are tilted out of the xy-plane just as all other bonds in the graphane region. The other extreme case would be that the interface bonds are completely in-plane, as if they would correspond to the graphene region. It is not known to which extent the bonds are tilted out of the plane. We believe that the situation is more like shown in Fig. \ref{fig_alpha_beta_interface}, because there should be a kind of a repulsive force between the hydrogen on the first graphane atom and the first graphene atom. {\it Ab initio} or structure calculations based on environment-dependent tight-binding models could yield a better estimate of the interface details.

In the following we quantify the tilting at the interface by a parameter $z_J$. The fully tilted situation (as shown in Fig. \ref{fig_alpha_beta_interface}) is described by setting $z_J=z_0$. The flat interface, on the other hand, is described by $z_J=0$.

\begin{figure}[!h]
\centering
\includegraphics[width=240pt]{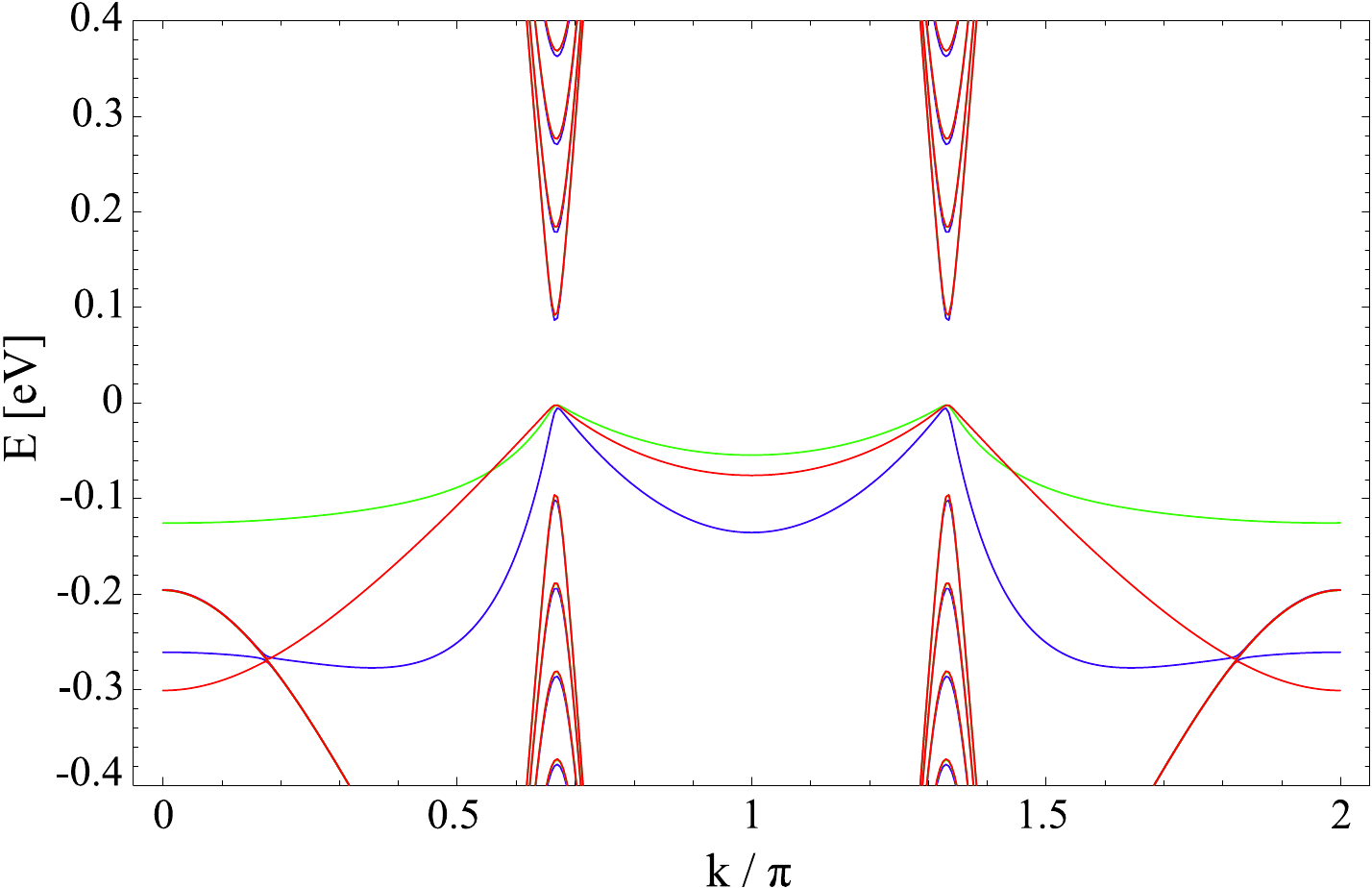}
\caption{Different band structures for $z_J=0$ (red), $z_J=z_0/2$ (green) and $z_J=z_0$ (blue). The graphene part of the $\alpha\beta$-ribbon is 21nm wide in this calculation and the graphane terminations are 2.1nm wide. The bulk states and the states at the outer graphane edges are not significantly affected by $z_J$. The corresponding energy bands lie on top of each other for the three cases.}
\label{fig_interface_detail_dispersion}
\end{figure}

The impact of the interface details on the dispersion of the GG-interface state is shown in Fig. \ref{fig_interface_detail_dispersion}. Obviously, primarily the edge states are affected by the interface details while the bulk dispersion is largely invariant. Also it is observed that, while the $\alpha$-edge state's dispersion always resembles a parabola very closely, the $\beta$-edge state changes the shape of its dispersion more heavily. However, for this work only the dispersion near K and K' is important. There, changing $z_J$ mainly renormalizes the interface state velocity by a factor of order unity.

In the numerical diagonalization of the $\alpha\beta$-ribbon Hamiltonian the dispersion of the $\beta$-edge states are sometimes (depending on the interface details) crossed by bands which are exponentially localized at the outer graphane edge (see Fig. \ref{fig_interface_detail_dispersion}). Usually, this crossing happens at $|k|\lesssim0.2\pi$. These states are spatially separated from the edge states at the interfaces and are thus not important here. If the graphane regions were infinitely wide, these bands would not affect the $\beta$-edge state at all. However, since the numerical calculations use a finite size ribbon, these graphane edge states have an exponentially small overlap with the $\beta$-edge state wave function. Since we are generally interested in the band structure near K and K', this crossing is not important for our reasoning. Thus, in the plots of the numerical spin splitting and the spin-orbit strength (see below), we leave out the part of the Brillouin zone which is beyond the crossing. This is why the plots of some numerical results are restricted to the $k$-range $[0.2\pi,1.8\pi]$.

\section{Effective spin-orbit Hamiltonian}

As explained above, the effective spin-orbit Hamiltonian of the edge states at the GG interface is obtained by projecting $H_{SO}$ (see Eq. (\ref{so_hamiltonian})) to the restricted Hilbert space of the edge states $\left|\psi^\alpha(k);\tau\right>$. This leads to an effective SOI described by
\begin{equation}
\Gamma_{\tau\tau'}(k) = \left<\psi^\alpha(k);\tau|H_{SO}|\psi^\alpha(k);\tau'\right>\label{gamma_form},
\end{equation}
which can be decomposed by means of spin Pauli matrices. It turns out that the function, multiplying $\sigma^x$ is zero while the coefficient functions of $\sigma^{y/z}$ are finite. This makes sense because in addition to the intrinsic SOI of pure graphene, there is a $z\rightarrow-z$ symmetry breaking at the interface which leads to a Rashba-like term $k_x\sigma^y$. The $x\rightarrow -x$ symmetry, however, is not broken by the interface so that the Dresselhaus-like term $k_x\sigma^x$ vanishes.

The function $\Gamma_{\tau\tau'}(k)$ can be calculated numerically from the wave functions of the GG edge states. However, it can well be approximated by an analytic form: the spin-orbit interaction at the GG interface is strong compared to the spin-orbit interaction in bulk graphene. Therefore, an effective model must respect the renormalization of the SOI by the amplitude of the interface state wave function directly at the interface. This is given by $\mathcal N_k^\alpha$ ($\mathcal N_k^\beta$) for $\alpha$- ($\beta$-) edge states in infinitely wide ribbons.

Moreover, because of time-reversal invariance, the $k$-dependent coupling strength must be odd at $k=0,\pi$. Thus, we only take into account odd polynomials around these two points. As it turns out, the linear terms are sufficient to fit the numerics of the $\alpha$-edge state. For the $\beta$-edge state more terms are needed for a quantitative fit.

\begin{figure}[!h]
\centering
\includegraphics[width=210pt]{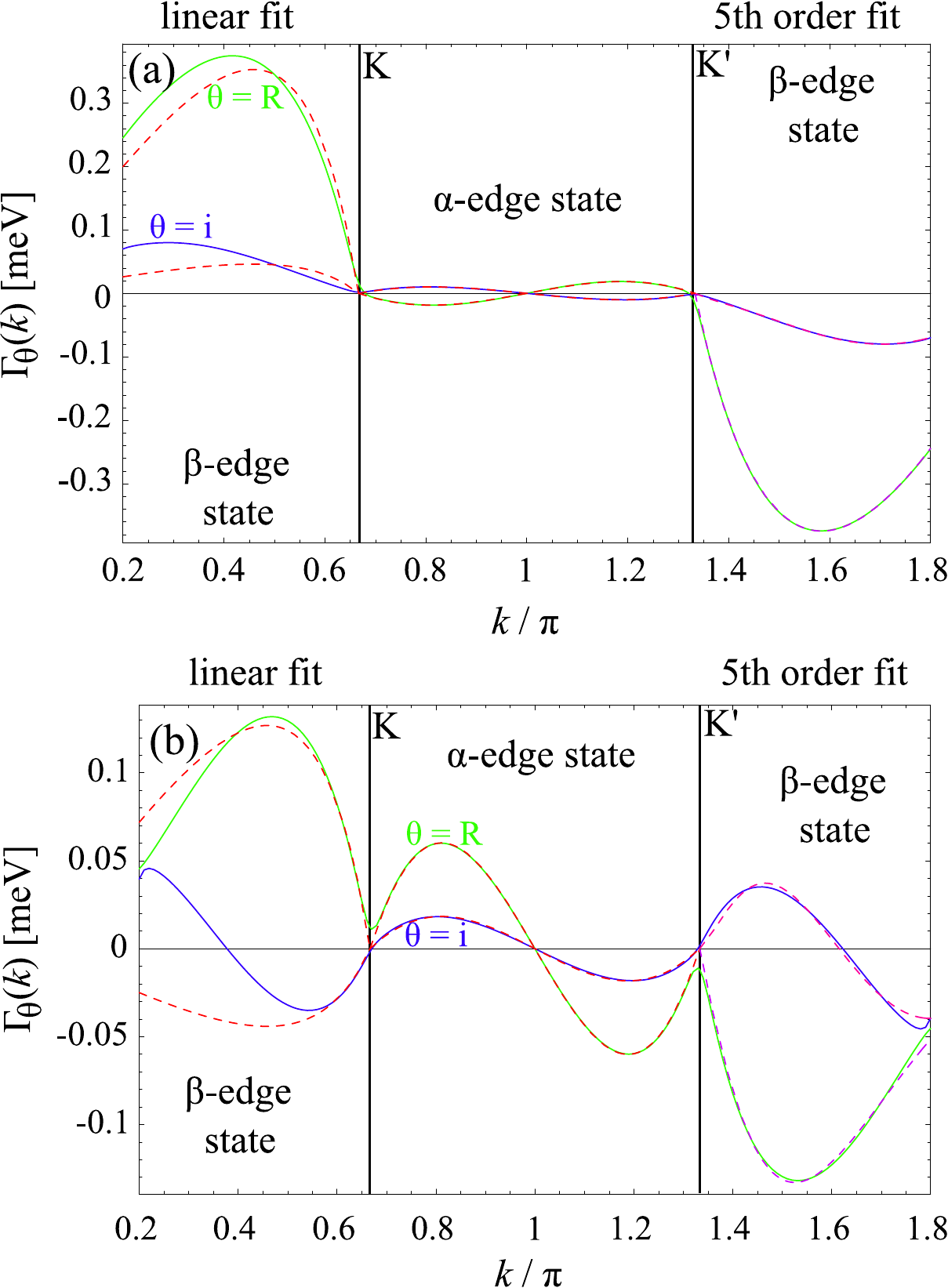}
\caption{Spin-orbit strength $\Gamma_\theta(k),\;\theta=i,R$ from the effective model (dashed lines) and from our extended tight-binding model (solid lines). The green (blue) line represents the Rashba (intrinsic) term of the spin-orbit Hamiltonian and the nearby dashed lines represent the corresponding terms from the effective models. The effective Hamiltonian of the $\alpha$-edge state is Eq. (\ref{eff_form_alpha}). The effective Hamiltonian of the $\beta$-edge state on the left-hand side (the region labeled by linear fit) is given by the linear version Eq. (\ref{eff_form_beta}) while on the right-hand side (the region labeled by 5th order fit), Eq. (\ref{eff_form_beta_3}) is shown. Part (a) shows a calculation for a flat interface ($z_J=0$) and part (b) shows a calculation for a maximally tilted interface ($z_J=z_0$).}
\label{fig_eff_hso_fit}
\end{figure}

The guessed form of the effective spin-orbit interaction of the $\alpha$- and $\beta$-edge states with only linear terms reads
\begin{eqnarray}
\Gamma_{SO,\alpha}^{\rm eff}(k) &=& (k-\pi)(\mathcal N^\alpha_k)^2\left[\Delta_R^\alpha \sigma^y + \Delta^\alpha_i \sigma^z\right ] \label{eff_form_alpha} \\
\Gamma_{SO,\beta}^{\rm eff} (k) &=& k (\mathcal N^\beta_k)^2\left[\Delta_R^\beta \sigma^y + \Delta^\beta_i \sigma^z\right ] . \label{eff_form_beta}
\end{eqnarray}
In order to find the parameters $\Delta_R^{\alpha/\beta}$ and $\Delta_i^{\alpha/\beta}$, we fit these functional forms to the numerical results. Fig. \ref{fig_eff_hso_fit} shows various fits. In Table \ref{tab_1_param_fit} the effective spin-orbit parameters for the flat ($z_J=0$) and the tilted ($z_J=z_0$) interfaces are given. The $\alpha$-edge state spin-orbit interaction is well reproduced by the effective form given in Eq. (\ref{eff_form_alpha}) while the $\beta$-edge state spin-orbit interaction is not. Therefore, we include higher powers in Eq. (\ref{eff_form_beta})
\begin{multline}
\Gamma^{\rm eff}_{SO,\beta} (k)= k(\mathcal N_k^\beta)^2 \left[\Delta_R^\beta (1+a_R^\beta k^2 + b_R^\beta k^4) \sigma^y \right.\\ +\left.\Delta^\beta_i (1+a_i^\beta k^2 + b_i^\beta k^4)\sigma^z \right]. \label{eff_form_beta_3}
\end{multline}
The parameters fit to the numerical results are given in Tab. \ref{tab_1_param_fit} for the one parameter forms (\ref{eff_form_alpha}) and (\ref{eff_form_beta}) and for the three parameter form (\ref{eff_form_beta_3}) in Tab. \ref{tab_3_param_fit}. In order to compare the effective models with the numerical results, we introduce the Rashba (intrinsic) spin-orbit strength $\Gamma_R(k)$ ($\Gamma_i(k)$) as the coefficient of the Pauli matrix $\sigma^y$ ($\sigma^z$) in the expressions (\ref{gamma_form}), (\ref{eff_form_alpha}), (\ref{eff_form_beta}), and (\ref{eff_form_beta_3}). Fig. \ref{fig_eff_hso_fit} compares the different expressions. Obviously, the linear expression (\ref{eff_form_alpha}) is sufficient for the $\alpha$-edge state while (\ref{eff_form_beta}) shows significant deviations from the numerical results for the $\beta$-edge state.

\begin{table}
\centering
\begin{tabular}{|c|c|c|} \hline
 & $z_j=0$ & $z_J=z_0$  \\ \hline
$\Delta_R^\alpha$ & -0.048 meV & -0.16 meV\\ \hline
$\Delta_i^\alpha$ & 0.026 meV & -0.050 meV \\ \hline
$\Delta_R^\beta$ & 0.44 meV & 0.13 meV\\ \hline
$\Delta_i^\beta$ &0.057 meV & -0.045 meV\\ \hline
\end{tabular}
\caption{Single parameter fit of the spin-orbit interaction for the edge states at $\alpha$- and $\beta$-interfaces.}
\label{tab_1_param_fit}
\end{table}

\begin{table}
\centering
\begin{tabular}{|c|c|c|} \hline
 & $z_j=0$ & $z_J=z_0$  \\ \hline
$\Delta_R^\beta$ & 0.56 meV & 0.056 meV\\
$a_R^\beta $ & -0.09 & 0.99\\
$b_R^\beta $ & $5\cdot 10^{-5}$ & -0.18 \\ \hline
$\Delta_i^\beta$ & 0.18 meV & 0.09 meV\\
$a_i^\beta $ & -0.40 & -0.89 \\
$b_i^\beta $ & 0.05 & 0.14\\ \hline
\end{tabular}
\caption{Three parameter fit of the spin-orbit interaction for the edge state at a $\beta$-interface.}
\label{tab_3_param_fit}
\end{table}

The spin splitting $\Delta\epsilon_{SO}(k)$ calculated from the effective model is in better agreement with the numerical results than the spin-orbit strengths $\Gamma_\theta(k)$ (see Fig. \ref{fig_spin_splitting}), even for the linear model of the $\beta$-edge state, because this quantity is an average over the intrinsic and Rashba term of the spin-orbit Hamiltonian.

For infinitely wide ribbons ($W\rightarrow\infty$), the normalization factors $\mathcal N_k^{\alpha/\beta}$ vanish at K,K'. Thus, the spin-orbit splitting at K,K' is not enhanced by the Hamiltonians (\ref{eff_form_alpha}-\ref{eff_form_beta_3}), compared to pure graphene. In finite size ribbons, however, the normalization of the edge state wave functions is significantly different from $\mathcal N_k^{\alpha/\beta}$ if $\xi^{\alpha/\beta}_k > W$. The edge state of a finite size $\alpha\beta$-ribbon with N unit cells in the y-direction can be written as
\begin{multline}
\left|\psi_0(k)\right> = \sqrt{\frac{1-|u_k|^2}{1-|u_k|^{2(N+1)}}} \times \\
\sum_{n=0}^N \exp\left(-\frac{n}{\xi_k^\alpha} + i n \phi\right) d_{n,k,B}^\dagger  \left|0\right>\label{analytical_wave_alpha_beta}.
\end{multline}
At K,K' the absolute value of this wave function is constant in y-direction. Thus, the amplitude at the interface is not proportional to $\left(\mathcal N^{\alpha/\beta}_{\rm K}\right)^2=0$ in this finite-size case, but rather proportional to
\begin{equation}
\frac{1-|u_k|^2}{1-|u_k|^{2(N+1)}} \xrightarrow{k\rightarrow {\rm K,K'}} \frac1{N+1}.
\end{equation}
As a result, the GG interface-induced SOI near K,K' is proportional to $W^{-1}$. Thus, the SOI of bulk graphene becomes dominant at K,K' in the limit $W\rightarrow\infty$.

\begin{figure}[!h]
\centering
\includegraphics[width=220pt]{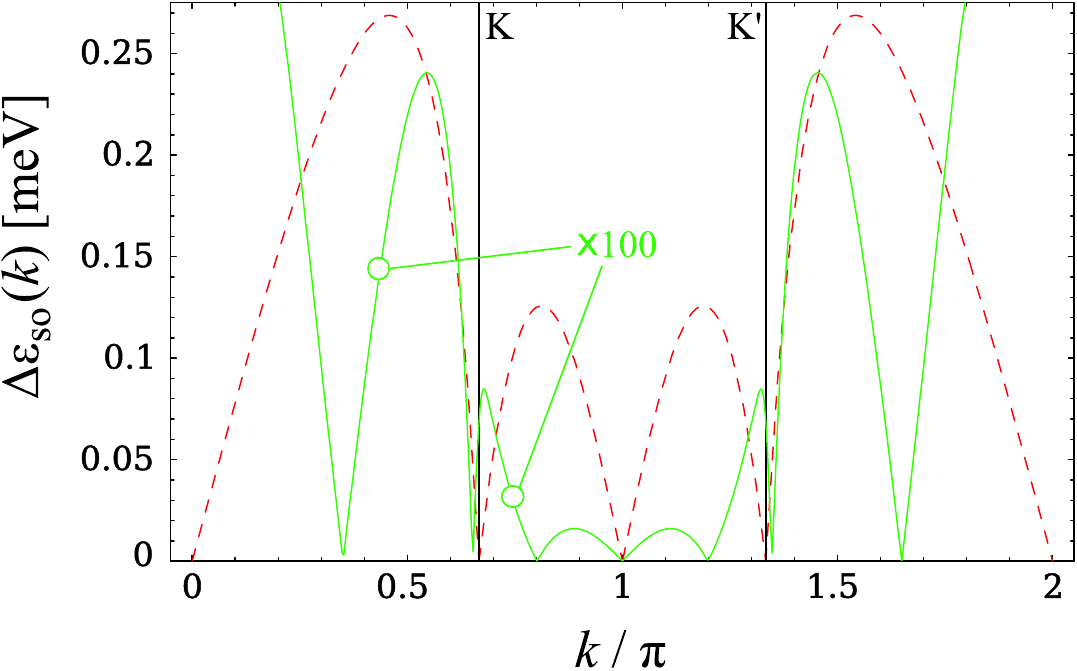}
\caption{Comparison between the interface-induced spin-orbit splitting calculated from the effective model (dashed red line) and the bulk graphene induced spin-orbit splitting (green line) the edge states. The bulk graphene induced spin-orbit splitting is scaled up by a factor of 100.}
\label{fig_split_comp}
\end{figure}

In Fig. \ref{fig_split_comp} we compare the spin-orbit splitting which is due to the GG interface, calculated from the effective models (\ref{eff_form_alpha}) and (\ref{eff_form_beta}), with the conventional spin-orbit splitting which is due to the bulk graphene SOI. Obviously, the magnitude of the interface-induced spin splitting is much larger than the conventional spin splitting for $k\neq$K,K'. However, at K,K' the interface-induced spin splitting vanishes because the edge states become delocalized over the whole graphene region so that the interface contribution to the spin splitting vanishes. This means that the spin splitting due to the bulk graphene SOI is dominant at K,K'.

This is important for the QSVHE in that the Fermi energy must be tuned exactly to the spin-orbit gap which is due to the bulk graphene SOI. Qualitatively, this bulk SOI can be included in the effective model by adding the bulk graphene-induced spin-orbit splitting of the edge states at K,K' \footnote{The bulk graphene-induced spin-orbit splitting of the edge states is $k$-dependent. However, it is sufficient to take only the value of the splitting at K,K'.}
\begin{equation}
H_{SO,\alpha/\beta}^{\rm eff} \rightarrow H_{SO,\alpha/\beta}^{\rm eff} + \Delta_0 \sigma^z \tau^3,\label{so_ext}
\end{equation}
where $2\Delta_0\simeq 1\mu$eV is the spin-orbit gap at K,K' in bulk graphene and $\tau^3=\pm1$ for K and K', respectively. The red, dashed curve in Fig. \ref{fig_split_comp}, calculated with the extended Hamiltonian (\ref{so_ext}) would differ in that it is not exactly zero at K,K' but $2\Delta_0$.


\begin{thebibliography}{99}
\bibitem{klein_footnote} Though, single dangling bonds have been observed experimentally (see Ref. \onlinecite{klein_edge}).
\end{thebibliography}
\end{document}